\documentstyle[frascatiphys,12pt]{article}
\begin{document}
\tolerance=10000
\hyphenation{has}
\hyphenation{that}

%\begin{figure}
%\epsfysize = 9in \epsffile {sokoloff_cover.ps}
%\end{figure}

\title{
\begin{flushright}
{\small IIT-HEP-95/1\\
hep-ex/9508015\\
January 1995\\
\vspace{.15 in}
}
\end{flushright}
PHYSICS OF AN ULTRAHIGH-STATISTICS CHARM EXPERIMENT\thanks{~~To
appear in
{\sl Proceedings of the HQ94 Workshop}, Univ. of Virginia, Oct.\ 7--10, 1994.}}
\author{Michael D. Sokoloff \\
{\em University of Cincinnati} \\
Daniel M. Kaplan\\
{\em Illinois Institute of Technology}}
\maketitle

\vspace{-.75 in}

\baselineskip=14.5pt
\begin{abstract}
We review the physics goals of an ultrahigh-statistics charm experiment and
place them in
the broader context of the community's efforts to study the Standard Model and
to search for physics beyond the Standard Model, and we point out some of the
experimental difficulties which must be overcome if these goals are to be met.

\end{abstract}
\baselineskip=17pt
%
%\section{--~ Introduction}

The CHARM2000 workshop\cite{CHARM2000} suggested that the goal
for a future experiment be a factor
$\sim10^2$ increase in statistics over the coming round of fixed-target
charm experiments at Fermilab (E781 and E831).
We consider the physics goals of such an ultrahigh-statistics
charm experiment ($ \sim $ 100 million reconstuctable $ D $'s ).
Some measurements will
test the Standard Model, some will measure its parameters,
and some will elucidate heavy-quark
phenomenology. We can outline the major goals as follows:

\begin{enumerate}

\item
Measurements which search for new physics\cite{Pakvasa}

\begin{itemize}
\parskip=0pt
\item $ D^0 - \overline D {}^0$ mixing
\item explicit flavor-changing neutral currents
\item direct {\it CP}  violation.
\end{itemize}

\item
Measurements which test the heavy-quark symmetry of QCD in the charm sector

\begin{itemize}
\parskip=0pt
\item form factors of semileptonic decays of charmed mesons
\item masses and widths of orbitally-excited charmed mesons.
\end{itemize}

These measurements are key to extracting
fundamental parameters from future beauty experiments.
\item
Measurements which probe aspects of perturbative and
nonperturbative QCD (including higher-twist effects)

\begin{itemize}
\parskip=0pt
\item nonleptonic singly and doubly Cabibbo-suppressed decay rates
\item dynamics of charm hadroproduction.
\end{itemize}
\end{enumerate}

\noindent
We next discuss these topics in more detail.

\section{--~ Searches for New Physics}

\subsection{\underline {$ D^0 - \overline D {}^0 $ mixing}}

$ D^0 - \overline D {}^0 $ mixing is one of the most interesting places to look
for physics beyond the Standard Model.
While Wolfenstein\cite{Wolfenstein}
once suggested large long-distance or dispersive
contributions to $ \Delta m_D $ within the Standard Model, more detailed
calculations by Donoghue {\it et
al.}\cite{Donoghue} give $ | \Delta m_D | \approx 10^{-6} $\,eV.
Recent analyses\cite{Georgi,Ohl} based on heavy-quark effective theory (HQET)
suggest
that
cancellations lead to $  | \Delta m_D | < 3.5 \times 10^{-8} $\,eV.
Values of $  | \Delta m_D | $ as large as  $  10^{-4} $\,eV are
possible in many models beyond the Standard
Model\cite{EHLQ}~$^-$\cite{HW}.
Thus there is a large window for observing new physics via $ D^0 -
\overline D{}^0 $ mixing.

A particularly intriguing example is discussed in a recent paper by Hall and
Weinberg\cite{HW}, which emphasizes that electroweak theories with several
scalar doublets are consistent with all known physics (especially with the
approximate {\it CP} symmetry in the neutral-kaon sector), provide an
alternative mechanism for {\it CP} violation,
and have various interesting phenomenological features.
In such models, tree-level scalar-exchange contributions to
neutral $ K $- and $ B $-meson mass mixing are at about the level
observed by experiments if Higgs bosons have masses in the 700\,GeV range.
Hall and Weinberg say that ``although this means that little can be
learned about the CKM matrix from $ \Delta m_K $ and $ \Delta m_B $, the
case of $ D - \overline D $ [mixing] presents different opportunities.
$ \ldots $
If we take the typical Higgs boson mass as near 1\,TeV to account for the
observed values of $ | \Delta m_K | $ and $ | \Delta m_B | $, then the
predicted value of $ | \Delta m_D | $ is close to the current experimental
limit, $ | \Delta m_D | < 1.3 \times 10^{-4} $\,eV."

CLEO II has reported a $ D^0 \rightarrow K^+ \pi^- $ signal with a
branching ratio about $ 2 \times \tan^4 \theta_C \times B ( D^0
\rightarrow K^- \pi^+) $\cite{Cinabro}.
This ``wrong-sign" kaon is a signature either of mixing or of doubly
Cabibbo-suppressed decay (DCSD). The two are
most easily separated in a fixed-target experiment, in which the lifetime
of the $ D $ can be directly measured: a DCSD signal decays exponentially,
while a mixing
signal has an additional $ t^2 $ dependence, so that it peaks at  $2\tau_D $.
If the CLEO II signal is primarily a DCSD signal, then it presents an
inescapable background for mixing studies using hadronic final states.
Assuming this is the case, an experiment such as we consider should be
sensitive to a mixing signal  on top of a DCSD
signal with $ | \Delta m_D |  \approx2 \times 10^{-5}  $\,eV.
Morrison\cite{Morrison} has pointed out that similar sensitivity might be
achievable also in semileptonic decays, which are free of the confounding
effects of DCSD. Liu's  thorough treatment\cite{Liu} includes the intriguing
suggestion that for mixing arising from the decay-rate difference between the
{\em CP} eigenstates $D_1$ and $D_2$, sensitivity an order of magnitude better
might be achievable in singly Cabibbo-suppressed modes, by using the
interference between mixing and DCSD to enhance the mixing signal.

\subsection{\underline {Charm-changing neutral currents}}

Some of
the models cited above\cite{Buchmuller,Joshipura} also allow the possibility of
explicit flavor-changing neutral currents (FCNC) in $ D $ decays.
E791 has reported\cite{Nguyen}
the best 90\%-confidence-level
upper limit for the branching ratio of $ D^+ \rightarrow \pi^+
\mu^+
\mu^- $ of $ 1.3 \times 10^{-5}$.
An ultrahigh-statistics experiment with good lepton identification
would have a sensitivity to this and other FCNC decays
(and also to lepton-number-violating decays)
one to two orders of magnitude lower.

\subsection{\underline {$C\!P$ violation}}

The Standard Model predicts that direct {\it CP} violation (observed as the
fractional difference between decay rates of particle and antiparticle to
charge-conjugate final states) will be of order $10^{-3}$ or less
in singly Cabibbo-suppressed $ D $ decays\cite{Bucella}.
(In the Standard Model, {\it CP} should be an exact symmetry for
Cabibbo-allowed and DCSD decays.)
Physics beyond the Standard Model might contribute {\it CP}-violating
amplitudes to decay rates, and there is a large window for observing new
physics. At the level of
statistics we consider here,
sensitivity to {\it CP} asymmetries at the fraction-of-a-percent level in
singly Cabibbo-suppressed decays and at the few-percent level in doubly
Cabibbo-suppressed decays may be possible.
Holding systematic uncertainties to the percent level will be challenging,
and experimenters planning to make such measurements must consider
carefully how systematic errors will be minimized and
how they will be measured.

\section{--~ HQET and Semileptonic Decays}

\subsection{\underline {Testing HQET via orbitally-excited charmed mesons}}

Within the Standard Model, it is generally agreed that heavy-quark symmetry
can be used to predict many nonperturbative properties of hadrons containing a
single heavy quark, and the most important of these predictions are for
exclusive semileptonic $ B $-meson decays\cite{Wisgur1}.
These nonperturbative effects will be
important in extracting $ V_{ub} $ and $ V_{cb} $ from measured decay rates.
Heavy-quark symmetry also relates the masses and widths of the
orbitally-excited
$ D^{**} $ mesons (including the $ D_s^{**} $ mesons), as has been discussed
recently in papers by Isgur and Wise\cite{Wisgur2},
Ming-Lu, Isgur, and Wise\cite{Wisgur3}, and
Eichten, Hill, and Quigg\cite{EHQ}.
While some authors argue that the charm-quark
mass is sufficiently large for the limit $ m_c  \rightarrow \infty $ to be a
good approximation\cite{EHQ},
others\cite{Godfrey} have argued that even for $ B $ mesons
the $ m_Q \rightarrow \infty $ limit has not been reached.
The experiment we consider will
measure the masses and widths of the orbitally-excited $ D^{**} $
mesons  with sufficient precision to confront theoretical models
quantitatively.
Where E691\cite{E691}, ARGUS\cite{ARGUS}, CLEO II\cite{CLEO}, and
E687\cite{E687} have measured the $
D_2^{*0} $, $ D_2^{*+} $, and $ D_1^0 $ widths with 50\% fractional errors,
such an experiment
should be able to achieve few-percent fractional errors.
To untangle the states and reflections which lie on top of each other,
it will also be necessary to measure $ \pi^0 $'s and (perhaps) single
photons well.
However, the benefit of making these measurements is that they
will establish how well heavy-quark symmetry works for charm and
give theorists the numbers they need to develop a more complete
phenomenology of $ B $ physics.

\subsection{\underline {Semileptonic form factors}}

High-statistics charm experiments will also contribute to our understanding of
the
form factors and helicity amplitudes of the vector mesons which can appear
as decay products in both $ D $ and $ B $ decays.
Extracting {\it CP}-violation parameters from measurements of
branching ratios for decays such as
$ B_d \rightarrow  \rho^0 \psi $ and
$ B_d \rightarrow K^* \psi $, which Dunietz\cite{Dunietz}
advocates as the best place to
measure the unitarity-triangle angle $  \gamma $,
requires the best possible measurement of the
$ \rho^0$ and $ K^* $ helicity amplitudes and form factors in the
$ D $ semileptonic decays
$ D^+ \rightarrow \rho^0 l \nu $ and $ D^+ \rightarrow K^{*0} l \nu $,
as they should be the same in $ D $ as in $B $ decay.
Assuming single-pole forms for the form factors, the mass of the pole
should be measurable  with better than 1\% precision.
In $ D^+
\rightarrow K^{*0} l \nu $,
it should be possible to
measure the polarization of the $ K^* $,
\begin{equation}
{\Gamma_L /
\Gamma_T} =
{{ \int P^*_V t | H_0(t) | ^2 dt}
\over
{\int P^*_Vt \left [ | H_+(t) |^2 + |H_- (t) |^2 \right ] dt} }
\end{equation}
(the ratio of
longitudinal to transverse form-factors),
with percent statistical and systematic uncertainties.
It should be possible to measure the polarization of
the $ \rho^0 $ in the Cabibbo-suppressed decay with  few-percent statistical
accuracy.
$ D_S \rightarrow \phi l \nu $ should  be measured with similar precision,
which will provide another test of the applicability of heavy-quark symmetry to
the study of semileptonic decays.

\subsection{\underline {Studying the CKM matrix with semileptonic decays}}

Studying semileptonic decays also contributes directly to our knowledge of
the CKM matrix.  High-statistics charm experiments are able to measure the
magnitudes of $ V_{cs} $ and $ V_{cd} $ from the semileptonic decays of the $ D
$ mesons. The absolute decay rates depend on various well-measured constants
(such as the $ D $ masses and lifetimes), the CKM matrix elements, and the
form factors of the hadrons produced along with the leptons.
Currently, $| V_{cd}| $ and $ | V_{cs} |$ are known with $ \pm 8 \% $ and $ \pm
20 \% $ precision\cite{PDG,Rosner}.
{}From the branching ratios for the semileptonic decays
$ D^0 \rightarrow \pi^- l^+ \nu_{l} $
and $ D^0 \rightarrow K^- l^+ \nu_{l} $,
the ratio $ |
V_{cd} | / | V_{cs} | $ should be determined
with a statistical accuracy of $\sim10^{-3}$.
In addition to testing the unitarity of the CKM matrix in the charm sector,
this ratio is
explicitly required to extract the unitarity angle $ \gamma $ from the ratio
$B( B_d \rightarrow  \rho^0 \psi )/B( B_d \rightarrow K^* \psi )$
discussed earlier.

\section{--~ Testing QCD with Charm Hadroproduction}

At the parton level, $ c \bar c $ production is supposed to be described by
perturbative QCD\cite{NDE}.
At the hadron level, the situation becomes more complicated.
Several experiments have reported large leading-particle effects at high
$x_F$\cite{E769}.
Leading-twist factorization in perturbative QCD predicts that
the charm quark's fragmentation is independent of the structure of the
projectile, while the data indicate that the produced charm quark coalesces or
recombines with the projectile spectator.
To test models of higher-twist effects\cite{Brodsky},
one wants to look at the observed production
asymmetries as functions of $  p_T $ and $ x_F $ jointly.
Measuring these asymmetries for different target nuclei
({\it i.e.} measuring the $ A $-dependence of these asymmetries)
will provide an extra handle on how quarks evolve into hadrons.

\section{--~ Experimental Issues}

Building an ultrahigh-statistics charm experiment will be a challenge.
The next-generation fixed-target experiments at Fermilab each project
reconstructed charm samples of order $ 10^6 $ events.
A Tau-Charm Factory operating at a luminosity of
$ 10^{33} \,$cm${}^{-2}$\,sec${}^{-1}$,
such as that proposed for SLAC\cite{slac343},
would reconstruct about $ 5 \times 10^6 $ charm per year.
The $ B $ factories planned for KEK and SLAC will produce of order $ 10^8 $
$ b \overline b $ events per year at design luminosity.
However, the number of reconstructed charm will be
similar to that projected for the Tau-Charm Factory.
HERA-$B$\cite{albrecht94} will produce a sufficient number of
$ D $'s  in $ p p $ collisions to imagine an ultrahigh-statistics experiment,
but the triggering requirements for charm physics
differ substantially from those for $ B $ physics,
and the data acquisition system is currently designed to operate at 10 Hz.
In addition, the current design for the HERA-$B$ vertex detector entails much
more multiple scattering and much poorer vertex resolution
than are desirable for a charm experiment.
There is no clear route to higher luminosities for $ e^+ e^- $ machines or
photon beams, so we are left with the problem of working with a relatively
small cross-section in a hadronic environment.
Whether it is a fixed-target experiment or a collider experiment that
we consider,
triggering selectively and efficiently will be the first major problem.

Building a detector which minimizes backgrounds will be another problem.
If we are looking for physics beyond the Standard Model, or looking for
relatively rare decays expected within the Standard Model, reducing backgrounds
will be at least as important as maintaining high efficiency for the
interesting signals. Two examples should suffice:
\begin{description}
\item [1)~ ]
The FCNC decay $ D^+ \rightarrow \pi^+ \mu^+ \mu^- $ is expected to have
a branching ratio less than $ 10^{-8} $ in the Standard Model\cite{schwartz93}
(except for the decay $ D^+ \rightarrow \phi \pi^+ $ followed by
$\phi \rightarrow \mu^+ \mu^- $, which populates a limited region of the
Dalitz plot).
E791\cite{Nguyen} finds that its sensitivity is greatest when the expected
number of background events is between 5 and 10 in the signal region.
If one were to scale up from this experiment simply, sensitivity would
increase only as the square root of the number
of reconstructed charm, since  the background would grow linearly
with the signal. To increase sensitivity here, it will be important to reduce
backgrounds without
substantially reducing efficiency for detecting muons.
This can be achieved by adding redundancy, {\it e.g.}  a second
view in the muon detector or a redundant muon-momentum measurement,
so that the double muon-misidentification probability becomes
approximately the square of the single muon-misidentification probability.
\item [2)~ ]
To measure the ratio of CKM matrix elements by comparing the decay
rates for $ D^+ \rightarrow K^* \ell \nu $ and $ D^+ \rightarrow \rho^0 \ell
\nu $, it will be critical to separate pions from kaons with a very high
degree of confidence;
the reflections of these signals feed into each other directly.
A fast RICH technology may suffice, but this is another area where
redundancy seems necessary to reduce the confusions which lead to
systematic errors.
\end{description}
Finally, it seems obvious that silicon pixel devices will
be necessary to provide both the
spatial resolution and the segmentation that are required for unambiguous
vertexing in the high-rate small-angle region.

\section{--~ Summary}

Charm physics provides a window into the Standard Model, and possibly beyond,
that complements those provided by other types of experiments.
In searches for $D-{\overline D}$ mixing,
FCNC, lepton-number-violating decays, or {\it CP}-violating
amplitudes, we are probing physics at the TeV level which may not
be accessible to other experiments: down-sector and up-sector
quarks can couple differently to new physics, and the charm quark is the only
up-sector quark for which such studies are possible.
Within the Standard Model, charm is probably the best place to test heavy-quark
symmetry quantitatively, and it is the best place to measure some of the
CKM matrix elements.
While ultrahigh-statistics experiments will be extremely difficult,
we can reasonably imagine that the technology will exist in the
next decade to reconstruct 100 million charm.
Getting from here to there will require a substantial R\&D effort,
and developing the expertise to design and build such an experiment
will require commitment from the individuals who will contribute directly,
from the laboratory at which it will be done, and from
the community as a whole.

We thank the organizers of this Workshop for offering us
this opportunity to discuss heavy-quark physics in such attractive
surroundings.
We also thank our colleagues from Fermilab proposal
P829\cite{p829}
for their contributions.
This work was supported by NSF grant PHY 92-04239 and
DOE grant DE-FG02-94ER40840.


\begin{thebibliography}{99}
\bibitem{CHARM2000}
D. M. Kaplan and S. Kwan, {\it eds.}, {\bf The Future of High-Sensitivity Charm
Experiments}, Proceedings of the CHARM2000 Workshop, Fermilab, June 7--9, 1994,
FERMILAB-Conf-94/190 (1994).

\bibitem{Pakvasa}
These topics have also been discussed recently by S. Pakvasa, ``Charm as
Probe of New Physics," in {\bf The Future of High-Sensitivity Charm
Experiments}, {\it op cit.}, p. 85.

\bibitem{Wolfenstein}
L. Wolfenstein, Phys.\ Lett.\ B {\bf 164}, 170 (1985).

\bibitem{Donoghue}
J. F. Donoghue, E. Golowich, B. R. Holstein and J. Trampetic,  Phys.\ Rev.\ D
{\bf 33}, 179 (1986).

\bibitem{Georgi}
Howard Georgi, Phys.\ Lett.\ B {\bf 297}, 353 (1993).

\bibitem{Ohl}
T. Ohl, G. Ricciardi, and E. H. Simmons, Nucl.\ Phys.\ {\bf B403}, 605 (1993).

\bibitem{EHLQ}
E. Eichten, I. Hinchliffe, K. D. Lane, and C. Quigg, Phys.\ Rev.\ D {\bf 34},
1547 (1986).

\bibitem{Bigi}
I. I. Bigi and A. F. Sanda, Phys.\ Lett.\ B {\bf 171}, 320 (1985).

\bibitem{Datta}
A. Datta, Phys.\ Lett.\ {\bf 154B}, 287 (1985).

\bibitem{Hadeed}
A. Hadeed and B. Holdom, Phys.\ Lett.\ {\bf 159B}, 379 (1985).

\bibitem{Buchmuller}
W. Buchmuller and D. Wyler, Phys.\ Lett.\ B {\bf 177}, 377 (1986);\hfil\break
Miriam Leurer, Phys.\ Rev.\ Lett.\ {\bf 71}, 1324 (1993).

\bibitem{Joshipura}
A. S. Joshipura, Phys.\ Rev.\ D {\bf 39}, 878 (1989).

\bibitem{HW}
Lawrence Hall and Steven Weinberg, Phys.\ Rev.\ D {\bf 48}, R979 (1993).

\bibitem{Cinabro}
D. Cinabro {\it et al.} (CLEO collaboration), Phys.\ Rev.\ Lett.\ {\bf 72},
1406 (1994).

\bibitem{Morrison}
R. Morrison, ``Charm2000 Workshop Summary,"
in  {\bf The Future of High-Sensitivity Charm
Experiments}, {\it op cit.}, p.\ 313.

\bibitem{Liu}
T. Liu, ``The $D^0 {\overline D}{}^0$ Mixing Search --- Current Status and
Future Prospects,"
in  {\bf The Future of High-Sensitivity Charm
Experiments}, {\it op cit.}, p.\ 375.

\bibitem{Nguyen}
A. Nguyen {\it et al.} (E791 Collaboration), ``Search for
$D^\pm \rightarrow \pi^+ \mu^+ \mu^-$," FERMILAB-Conf-94/187-E,
contributed to the 27th International Conference on High Energy Physics
(ICHEP), Glasgow, Scotland, 1994.

\bibitem{Bucella}
F. Bucella {\it et al.}, Phys.\ Lett.\  B {\bf 302}, 319 (1993).

\bibitem{Wisgur1}
N. Isgur and M. B. Wise, Phys.\ Lett.\ B {\bf 232}, 113 (1989) and
Nucl.\ Phys.\ {\bf B348}, 276 (1991).\hfil\break
Howard Georgi, Nucl. Phys. {\bf B348}, 293 (1991).

\bibitem{Wisgur2}
N. Isgur and M. B. Wise, Phys. Rev. Lett. {\bf 66}, 1130 (1991).

\bibitem{Wisgur3}
Ming-Lu, M. B. Wise, and N. Isgur, Phys. Rev. D {\bf 45}, 1553 (1992).

\bibitem{EHQ}
E. Eichten, C. T. Hill and C. Quigg, Phys. Rev. Lett. {\bf 71}, 4116 (1993);
``Spectra of Heavy-Light Mesons," in {\bf The Future of High-Sensitivity
Charm Experiments}, {\it op cit.}, p.\ 345, and ``Orbitally Excited Heavy-Light
Mesons Revisited," {\it ibid.}, p.\ 355.

\bibitem{Godfrey}
S. Godfrey and R. Kokoski, Phys. Rev. D {\bf 43}, 1679 (1991).

\bibitem{E691}
J. C.
Anjos {\it et al.} (E691 Collaboration), Phys.\ Rev.\ Lett.\ {\bf 62}, 1717
(1989).

\bibitem{ARGUS}
H. Albrecht {\it et al.} (ARGUS Collaboration), Phys.\ Lett.\ B {\bf 221}, 422
(1989);
{\it ibid.} {\bf 231}, 208
(1989); and
{\it ibid.} {\bf 232}, 398
(1989).

\bibitem{CLEO}
P. Avery {\it et al.} (CLEO Collaboration), Phys.\ Rev.\ D {\bf 41}, 774
(1990).

\bibitem{E687}
P. L. Frabetti {\it et al.} (E687 Collaboration), Phys.\ Rev.\ Lett.\ {\bf 72},
324 (1994).

\bibitem{Dunietz}
Isard Dunietz, ``{\it CP} Violation with Additional $ B $ Decays," in  ``{\bf
$ B $ Decays}," Sheldon Stone, {\it ed.}, World Scientific, Singapore
(1992), p.\ 393, and
``Extracting CKM Parameters from $ B $
Decays," FERMILAB-Conf-93/90-TH, in {\bf Proceedings of the Workshop on B
Physics at Hadron Accelerators}, Snowmass, CO, June 21 -- July 2, 1993,
P.~McBride and C.~S.~Mishra, {\it eds.}, FERMILAB-Conf-93/267 (1993), p.\ 83.

\bibitem{PDG}
Particle Data Group, L. Montanet {\it et al.}, Phys.\ Rev.\ D {\bf 50}, Part 1
(1994).

\bibitem{Rosner}
J. L. Rosner, ``The Cabibbo-Kobayashi-Maskawa Matrix,"
in  ``{\bf
$ B $ Decays}," Sheldon Stone, {\it ed.}, World Scientific, Singapore
(1992), p.\ 312.

\bibitem{NDE}
P. Nason, S. Dawson and R. K. Ellis, Nucl.\ Phys.\ {\bf B327}, 49 (1989) and
ERRATUM,  {\it ibid.}, {\bf B335}, 260 (1990).

\bibitem{E769}
G. A. Alves {\it et al.} (E769 Collaboration), Phys.\ Rev.\ Lett.\ {\bf 72},
812 (1994);\hfil\break
T. Carter {\it et al.} (E791 Collaboration), ``Production Asymmetries in $x_F$
and $p_T^2$ for $D^{\pm}$ Mesons," FERMILAB-Conf-94-185-E,"
contributed to the 27th International Conference on High Energy Physics
(ICHEP), Glasgow, Scotland, 1994.

\bibitem{Brodsky}
S. J. Brodsky, J. F. Gunion, and D. E. Soper, Phys.\ Rev.\ D {\bf 36}, 2710
(1987);\hfil\break
S. J. Brodsky, P. Hoyer, A. H. Mueller, and W.-K. Tang, Nucl.\ Phys.\ {\bf
B369}, 519 (1992);\hfil\break
W.-K. Tang, ``Anomalous Charm Production at Large $x_F$,"
in {\bf The Future of High-Sensitivity Charm Experiments}, {\it op cit.}, p.\
251.

\bibitem{slac343}
{\bf Proceedings of the Tau-Charm Factory Workshop},
L. V. Beers, {\it ed.}, SLAC-343, May 1989.

\bibitem{albrecht94}
W. Hofmann and W. Schmidt-Parzefall, spokespersons,
``HERA-$B$: An Experiment to Study CP
Violation in the $B$ System Using an Internal Target at the HERA Proton Ring,"
DESY-PRC 94/02 (1994); also
H. Albrecht, these Proceedings.

\bibitem{schwartz93}
A. Schwartz, Mod. Phys. Lett. {\bf A8}, 967 (1993).

\bibitem{p829}
B. T. Meadows {\it et al.}, ``P829: Continued Study of Heavy Flavors at TPL,"
a proposal to the Fermi National Accelerator Laboratory, November 1993, and
addendum, February 1994.

\end{thebibliography}
\end{document}